\documentclass[a4paper,fleqn,usenatbib]{mnras}
\usepackage{newtxtext,newtxmath}

\usepackage[T1]{fontenc}
\usepackage{ae,aecompl}

\usepackage{graphicx}	
\usepackage{amsmath}	
\usepackage{amssymb}	
\usepackage{enumerate}

\title[Nonlinear difference between images of Q0957+561]{Revealing the nonlinear behaviour of the lensed quasar Q0957+561}

\author[A. Bewketu et al.]{
A. Bewketu Belete,$^{1}$\thanks{E-mail: asnakew@fisica.ufrn.br}
B. L. Canto Martins,$^{1}$
I. C. Le\~ao,$^{1}$
 J. R. De Medeiros$^{1}$
\\
$^{1}$Departamento de F\'isica Te\'orica e Experimental, Universidade Federal do Rio Grande do Norte, Natal, RN 59078-970, Brazil\\
}

\date{Accepted 2018 December 10. Received 2018 December 10; in original form 2018 October 3.}

\pubyear{2018}

\begin{document}
\label{firstpage}
\pagerange{\pageref{firstpage}--\pageref{lastpage}}
\maketitle

\begin{abstract}
Knowledge about how the nonlinear behaviour of the intrinsic signal from lensed background sources changes on its path to the observer provides much information, particularly about the matter distribution in lensing galaxies and the physical properties of the current universe, in general. Here, we analyse the multifractal (nonlinear) behaviour of the optical observations of A and B images of Q0957+561 in the $r$ and $g$ bands. AIMS: To verify the presence, or absence, of extrinsic variations in the observed signals of the quasar images and investigate whether extrinsic variations affect the multifractal behaviour of their intrinsic signals.
METHOD: We apply a wavelet transform modulus maxima-based multifractality analysis approach. 
RESULTS: We detect strong multifractal (nonlinear) signatures in the light curves of the quasar images. The degree of multifractality for both images in the $r$ band changes over time in a non-monotonic way, possibly indicating the presence of extrinsic variabilities in the light curves of the images, i.e., the signals of the quasar images are a combination of both intrinsic and extrinsic signals. Additionally, in the r band, in periods of quiescent microlensing activity, we find that the degree of multifractality (nonlinearity) of image A is stronger than that of B, while B has a larger multifractal strength in recent epochs (from day 5564 to day 7527) when it appears to be affected by microlensing. Finally, comparing the optical bands in a period of quiescent microlensing activity, we find that the degree of multifractality is stronger in the $r$ band for both quasar images. In the absence of microlensing, the observed excesses of nonlinearity are most likely generated when the broad-line region (BLR) reprocesses the radiation from the compact sources.
\end{abstract}




\section{Introduction} \label{sec:intro}
Though quasars, in general, are known by their extremely high luminosities, gravitationally lensed quasars (hereinafter GLQs) are brighter than their unlensed counterparts and produce different image components \citet{2015MNRAS.448.1446A}. Study of GLQs provides significant information, mainly about background source quasar variability mechanisms and accretion disk structure \citep{2015ApJ...799..149J, 2007ApJ...661...19P}, the mass distribution in lensing galaxies \citep{2008MNRAS.391.1955B} and, in general, information to constrain physical properties of the Universe \citep{2018A&A...616A.118G, 2018MNRAS.476..663K}. It has been understood that the time delay caused by gravitational lensing is directly related to the current expansion rate of the Universe (the Hubble constant) and the mean surface mass density of the lensing galaxy (e.g., \citealt{2004mmu..symp..117K, 1964MNRAS.128..307R}). In the study of GLQs, great attention has been devoted to determining time delays between images, constraining the Hubble constant, disentangling intrinsic and extrinsic signals and identifying their variability mechanisms.\\
The quasar Q0957+561, at z = 1.41, is the first identified GLQ \citep{walsh19790957+} and one of the most studied lensed quasars (e.g., \citealt{hainline2011new, nakajima2009improved, cuevas2006accurate, schild1996microlensing, rhee1991estimate, gondhalekar1980uv}); its lensing galaxy, which is known to be part of a cluster of galaxies, is located at z = 0.36 \citep{keeton2000host, rhee1991estimate}. The doubly imaged quasar Q0957+561 has been observed since its discovery in 1979 \citep{walsh19790957+} in different electromagnetic bands, including the $r$ (read arm) and $g$ (blue arm) bands, both within the optical ranges. It has been said that lensing is due to matter inhomogeneity at some point between the source and observer, and intervening matter (mainly lensing galaxies) is known to split light from background sources \citep{wambsganss1998gravitational}. The travel times between image components of background source do not agree with each other due to the difference in their ray paths. As a result, this difference introduces a time delay between components of the same source, which in turn provides extremely valuable astrophysical information.\\
\citet{shalyapin2008new} studied the light curves of Q0957+561 in the $g$ and $r$ passbands, determined a time delay $\Delta t_{BA}$ = 417 $\pm$ 2 d in the g band (1$\sigma$, A is leading), and used data for image A to measure the delay between a large event in the g band and the corresponding event in the r band. The time delay between both optical bands was $\Delta t_{rg}$ = 4 $\pm$ 2 d (1$\sigma$, g band event is leading).  In the same paper, they demonstrated the absence of extrinsic variability in the quasar light curves and identified that reverberation within the gas disc around the supermassive black hole is a possible mechanism for the observed intrinsic variability (see also \citealt{2012ApJ...744...47G}). From other monitoring data of Q0957+561 in several optical bands, time delays between the two images A and B are found to be mainly in the range from 417-425 d (A is leading; e.g., \citealt{2003ApJ...587...71C, 2003A&A...402..891O, 0004-637X-552-1-81}). A more recent study of Q0957+561 has shown that $\Delta t_{rg} \sim$ 1 d for image A and $\Delta t_{rg} \sim$ 4 d for image B, which agrees with $\Delta t_{BA} \sim$ 420 d in the $r$ band \citep{shalyapin20125}. This extra delay of approximately 3 d in the $r$ band was also favoured in a previous study of the lens system \citep{1997ApJ...482...75K} and is roughly consistent with several estimates in red bands (e.g., \citealt{2003A&A...402..891O, 1999ApJ...526...40S}). Though \citet{shalyapin20125} considered a dense cloud within the cD lensing galaxy along the line of sight to the image A as a possible cause for a 3 d lag between optical bands, this lag relies on standard cross-correlation techniques that could lead to biased results, so the true time delay $\Delta t_{BA}$ is most likely achromatic \citep{2018A&A...616A.118G}.   Even though no microlensing effects have been detected in the light curves of the lensed quasar Q0957+561 over several decades, strong evidence for extrinsic variability was found in the initial and last years of monitoring, as reported in \citet{1998A&A...336..829P} and \citet{2018A&A...616A.118G}.\\
Here, we study the multifractal (nonlinear) behaviour of the optical observations of the lensed quasar Q0957+561 in the $r$ and $g$ bands separately. Our questions are the following: Is there any difference in nonlinearity between the signals of the images A and B of Q0957+561? What can we learn about the intrinsic variability of the quasar and the extrinsic mechanisms distorting it? To address these questions, we analyse the multifractal (nonlinear) behaviour of the light curves of images A and B of the gravitationally lensed quasar Q0957+561 in the $r$ and $g$ bands using a wavelet transform-based multifractality analysis approach called Wavelet Transform Modulus Maxima (hereinafter WTMM). 
\citet{doi:10.1093/mnras/sty1316} have shown that radio quasars are intrinsically multifractal and that the multifractality (or nonlinearity) strength is different across electromagnetic spectrum. In addition, \citet{2018Belete} verified that redshift correction does not affect quasars' multifractality behaviour, which is the same in both the observed and rest frames of quasars. Multifractality analysis has been applied in different science cases (e.g., \citealt{2017AdSpR..60.1363M, 2016JAsGe...5..301M, 2016cosp...41E.946K, 2016AGUFM.P11A1846A, 2016ApJ...831...87D, 2013EPSC....8...30A, 2012SPIE.8222E..0FJ, 2011Ouahabi, 2009PhPl...16j2307N, 2007EPJB...60..483L, 1998adap.org..8004D}).\\
Our work is organized as follows. In Section \ref{data}, we present the light curves, method, and procedures. We discuss the results in Section \ref{res}. The summary and conclusions are included in Section \ref{concl}.
\section{Data Collection, Method and Procedures}\label{data} \vspace{0.1pt}
\subsection{Data collection: Light curves}
\begin{figure*}
\centering
\includegraphics[scale=0.4]{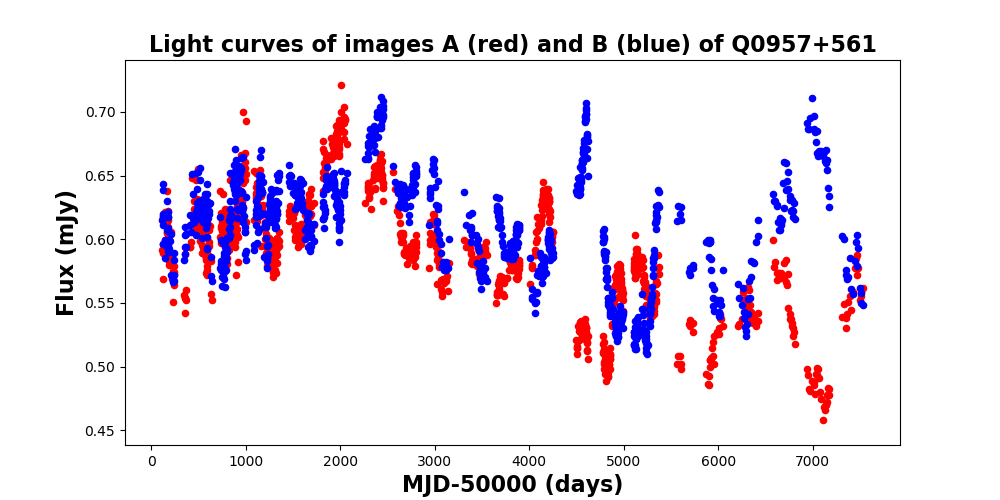} 
\includegraphics[scale=0.4]{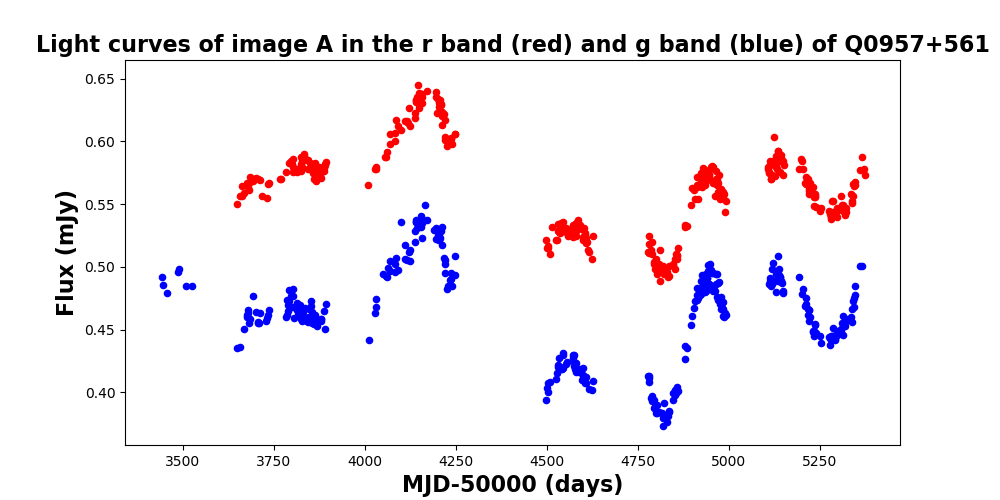}
\includegraphics[scale=0.4]{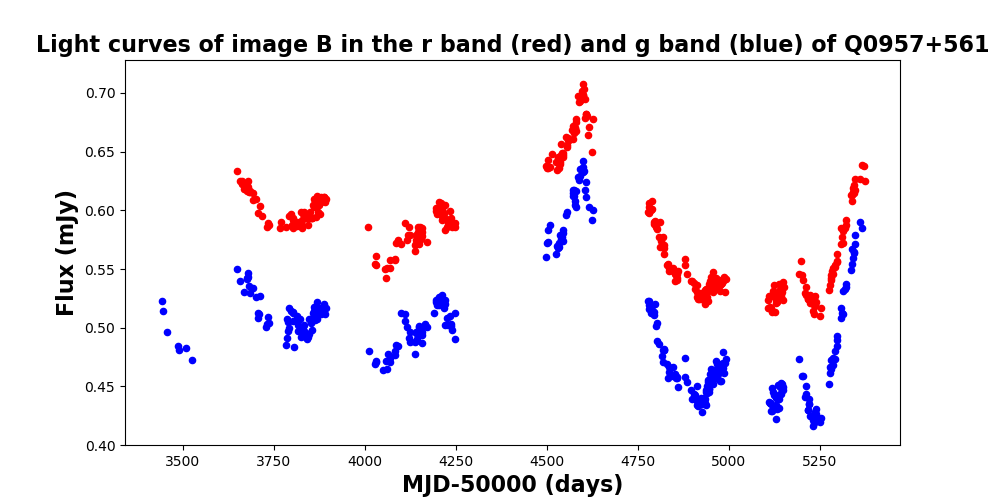}
\caption{Top: Light curves of images A and B -- red and blue, respectively -- of Q0957+561. Middle: Light curves of image A in the $r$ band (red) and $g$ band (blue) of Q0957+561. Bottom: Light curves of image B in the $r$ band (red) and $g$ band (blue) of Q0957+561.}
\label{fig1}
\end{figure*}

We use the the long-term optical light curves of Q0957+561 (1996-2016) for images A and B in the $r$ band (Fig. \ref{fig1}, top panel) and the 5.5-year (2007-2010) optical observational data of Q0957+561 in the $r$ and $g$ bands (Fig.\ref{fig1}, middle and bottom panels, respectively) from the GLQ database, which can be accessed via  the 
GLENDAMA website\footnote{http://grupos.unican.es/glendama/database/}. In the results section of this website, the long-term light curves in the $r$ band (those spanning from 1996 to 2016) are given in magnitudes. Since our aim is to analyse multifractal signatures in the flux observations (if any), we have converted the magnitudes of the signals to their equivalent fluxes according to the relation:  
\begin{equation}\label{eq1}
Flux =(3631 Jy) \times 10 ^{(-0.4 \times magnitude)}
\end{equation}
where $Jy$ stands for Jansky.  For more information about the light curves used in this study, refer to \citet{2018A&A...616A.118G}, \citet{shalyapin20125}, and \citet{2008NewA...13..182G}. 
\subsection{Method and Procedures}
The continuous wavelet transform is an excellent tool for mapping the changing properties of nonstationary signals. Because of its capability of decomposing a signal into small fractions that are well localized in time and frequency and detecting local irregularities of a signal (areas of the signal where a particular derivative is not continuous) such as nonstationarity, oscillatory behaviour, breakdown, discontinuity in higher derivatives, the presence of long-range dependence, and other trends \citep{maruyama2016solar, puckovs2012wavelet}, wavelet analysis remains one of the most preferred signal analysis techniques to date. Additionally, there is a claim that wavelet transforms are suitable for multifractal analysis and allow reliable multifractal analysis to be performed \citep{muzy1991wavelets}. Here, we apply WTMM-based multifractality analysis, which was originally introduced by \citet{muzy1991wavelets}. We follow the procedures discussed by \citet{puckovs2012wavelet}:  \\
1. We calculate wavelet coefficients of the signal $X$($t$) using the following mathematical relation:
\begin{equation}\label{eq2}
W(s,a)=  \frac{1}{\sqrt{s}}\int_{0}^{T}\left(\Psi\left(\frac{t-a}{s}\right). X(t)\right)dt
\end{equation}
where $W$ is the wavelet coefficients; $\Psi$($s$,$a$,$t$) is the mother wavelet function;
$s$ is the scaling parameter;
$a$ is the shift parameter;
$X$ is the signal;
$t$ is the time at which the signal is recorded; and
$T$ is maximal time value or signal length.
The analysing  wavelet $\Psi$($t$) is generally chosen to be well localized
in space and frequency. At lower scales $s \approx 0$, the number of local maxima lines (hereinafter LcMx) tends to infinity. Though it has been suggested that the scaling parameter $s$ used in the WTMM approach should be limited to $s \leq$ [1, 28], it should be in the interval [1, $\frac{T}{2}]$ and also can be in the interval [1, $\frac{T}{4}$], mainly to reduce the computation time. Since it is the maxima line that points towards regularities or carries information about any singularity or nonlinearity in the signal $X$($t$), we only calculate wavelet coefficients that correspond to local maxima lines; therefore, it is unnecessary to calculate wavelet coefficients that do not contain maxima lines. Hence, determining the limit of the most informative maximal scales is dependent on the presence and absence of local maxima lines. In our case, knowing that it is not important to calculate wavelet coefficients at large scales that do not correspond to maxima lines (rather, it is only a waste of time), our scale parameter is determined to be in the interval [1, $\frac{T}{4}$], which is still informative \citep{puckovs2012wavelet}. The shifting parameter $a$ cannot be greater than the signal length $T$, and therefore, $a$  $\leq$ T.\\
Usually, $\Psi$($t$) is only required to be of zero mean, but in addition to these requirements, for the particular purpose of multifractal analysis, $\Psi$($t$) is also required to be orthogonal to some low-order polynomials, up to the degree $n$-1 (i.e., to have $n$ vanishing
moments) \citet{2006GeoJI.164...63E}:
\begin{equation}\label{eq3}
\int_{-\infty}^{+\infty} t^{m}\Psi(t)dt= 0,      \forall m, 0 \leq m \le n
\end{equation} 
The Mexican Hat (MHAT) wavelet (second-order Gauss wavelet) has been chosen to be the analysing wavelet. This wavelet is one of the wavelets that has been applied for WTMM-based analysis and is represented by the relation
\begin{equation}\label{eq4}
\Psi(t)= (1-t^{2}).e^{-\frac{t^{2}} {2}}
\end{equation} 
where $\Psi$($t$) is the analysing wavelet function; $t$ is the time at which the signal is recorded. \\
2. We represent the calculated absolute wavelet coefficients in matrix form as
\begin{equation}\label{eq5}
W_{s,a}^{sq}=(W(s,a))^{2}|(s,a\in N)\wedge(s \in [1, s_{max}])\wedge(a \in [1, T]),
\end{equation}
where $W^{sq}$ is the squared wavelet coefficients matrix; $s_{max}$ is the maximal scaling parameter;
$s$ is the scaling parameter; $a$ is the shifting parameter; and $T$ is the signal length. \\
3. We calculate the skeleton function from the squared wavelet coefficient matrix and express it in matrix form as 
\begin{equation}\label{eq6}
LcMx_{s,a}=LcMx(s,a)
\end{equation}
under the conditions ($s$,$a\in N$) $\wedge$ ($s \in [1, s_{max}$]) $\wedge$ ($a \in [1, T]$), where $LcMx$ is the wavelet skeleton function; $W$($s$, $a$) are the wavelet coefficients;
$s$ is the scaling parameter; $s_{max}$ is the maximal scaling parameter; $a$ is the shift parameter; and $T$ is the signal length.
The skeleton function is a collection of local maxima lines at each scale, i.e., it is the scope of all local maxima lines that exist on each scale $s$. In other word, skeleton matrix construction is a technique of excluding coefficients from the absolute wavelet coefficients matrix that are not maximal. As a result, in the skeleton matrix, only absolute wavelet coefficients that correspond to local maxima lines exist. The need to collect all the maxima lines at each scale together in matrix form, the skeleton function, is because it is the maxima lines that carry valuable information about the signals, i.e., maxima lines point towards regularities in the signal. Since wavelet coefficients on corners provide minimal or no information, and consequently local maxima lines (LML) on corners also provide no significant information about the singularity in the signal, we therefore take the edge effect into consideration by removing the local maxima lines on corners using the formula provided by \citet{puckovs2012wavelet}. We fix broken lines, gaps, and single points in the LcMx matrix by applying an algorithm called supremum algorithm, which consists of seven steps, as explained by \citet{puckovs2012wavelet}.\\
4. Using the collected local maxima lines, we calculate the thermodynamic partition function, a function that connects the wavelet transform and multifractality analysis part as follows:
\begin{equation}\label{eq7}
Zq(s)=\sum_{a=1}^{T-1}(C(s).WTMM)^{q}|(LcMx_{s,a}=1),
\end{equation} 
where $Zq(s)$ is the thermodynamic partition function;
$WTMM$ is the wavelet modulus maxima coefficients;
$C$($s$) is a constant depending on scaling parameter $s$;
$s$ is the scaling parameter;
$q$ is the moment, which takes any interval with zero mean -- in our case, $q$ $\in$ [-5, 5]; and
$LcMx$ is the wavelet skeleton function (aggregate of local maxima lines in matrix form).
 The thermodynamic partition function is a function of two arguments - the scaling parameter $s$ and power argument $q$. The moment $q$ discovers different regions of singularity measurement in the signal, i.e., it indicates the presence of wavelet modulus maxima coefficients of different values. The condition $LcMx_{s,a}$ = 1 is to inform that only modulus maxima coefficients are used. 
What is important here is the relationship between $Zq$($s$) and $s$, which determines the scalability of the signal under consideration.  In the WTMM approach, the wavelet transform maxima are used to define a partition function, whose power-law behaviour is used for an estimation of the local exponents.  On small scales, the following relation is expected: 
\begin{equation}\label{eq8}
Zq(s)\sim s^{\tau(q)},
\end{equation}
where $\tau(q)$ is the scaling exponent function, which is the slope of the linear fitted line on the log-log plot of $Zq$($s$) and  $s$ for each $q$.\\
5. We determine the scaling exponent function $\tau$($q$) using the following relation:  
\begin{equation}\label{eq9}
\tau(q)= \lim_{s\to 0}\frac{\ln(Zq(s))}{ln(a)},
\end{equation}
where $Zq$($s$) is the thermodynamic partition function;
$\tau$ is the local scaling exponent;
$s$ is the scaling parameter; and
$q$ is the moment.
The condition $\tau$($q$ = 0) $+$ 1 = 0 is important for multifractal spectrum calculation.
The scaling exponent function $\tau$($q$) is a function of one argument $q$ and is determined from the slope of the  line fitted line on the log-log plot of $Zq$($s$) against the logarithm of time scale $s$ for each $q$, which means that the behaviour of the scaling function $\tau$($q$) is completely dependent on the nature of the thermodynamic partition function.
We define monofractal and multifractal as follows: a time series is said to be monofractal  if $\tau$($q$) is linear with respect to $q$; if $\tau$($q$) is nonlinear with respect to $q$, then the time series considered is classified as multifractal (Frish and Parisi, 1985).\\
6. At last, once we determine the scaling exponent $\tau$($q$), it is necessary to estimate the multifractal spectrum $f$($\alpha$) to be able to fully draw conclusions about the multifractal or nonlinear behaviour of a considered signal. We estimate the multifractal spectrum function  $f$($\alpha$) via the Legendre transformation as follows \citet{halsey1986fractal}:
\begin{equation}\label{eq10}
\alpha = \alpha(q) = \frac{\partial \tau(q)}{\partial q},
\end{equation}
where $\alpha$ is the singularity exponent or Holder exponent.
\begin{equation}\label{eq11}
f(\alpha) = q.\alpha - \tau(q),
\end{equation}
where $f$($\alpha$) is the multifractal spectrum function. 
Smaller values of $\Delta\alpha$ (i.e., $\Delta\alpha$ nears zero) indicate the monofractal limit, whereas larger values indicate the strength of the multifractal behaviour in the signal \citep{telesca2004investigating, ashkenazy2003nonlinearity, shimizu2002multifractal}.  When the multifractal structure is sensitive to the small-scale fluctuation with large magnitudes, the spectrum will be found with right truncation, whereas the multifractal spectrum will be found with left-side truncation when the time series has a multifractal structure that is sensitive to the local fluctuations with small magnitudes. 
\section{Results and Discussion}\label{res}
Here we analyse, and discuss, the multifractal behaviour of the light curves given in Fig. \ref{fig1}.  By using  Eqs. \ref{eq1} through \ref{eq3}, we compute the absolute wavelet coefficients and construct the corresponding skeleton functions by collecting absolute wavelet coefficients that only hold local maxima lines for all the light curves considered. The multifractality analysis for each light curve in Fig. \ref{fig1} is discussed below.
\subsection{Analysis of the light curves of images A and B in the $r$ band}\label{subsec1}
\begin{figure*}
\centering
\includegraphics[scale=0.39]{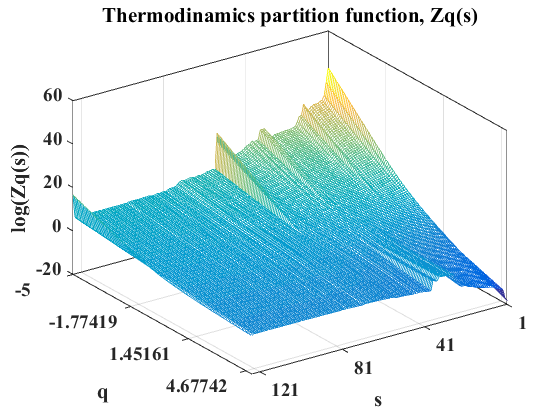} 
\includegraphics[scale=0.39]{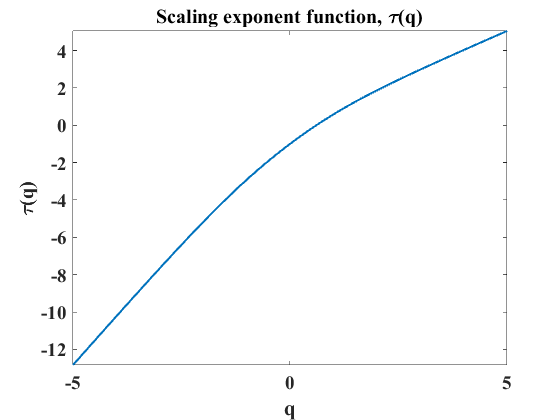}
\includegraphics[scale=0.39]{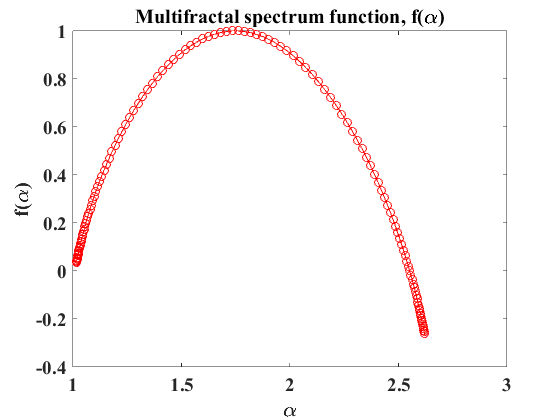}
\caption{The thermodynamic partition function $Zq$($s$) (left), the scaling exponent function $\tau$($q$) (middle), and the multifractal spectrum function $f$($\alpha$) (right) of image A of Q0957+561 in the $r$ band.}
\label{fig2}
\end{figure*}

\begin{figure*}
\centering
\includegraphics[scale=0.39]{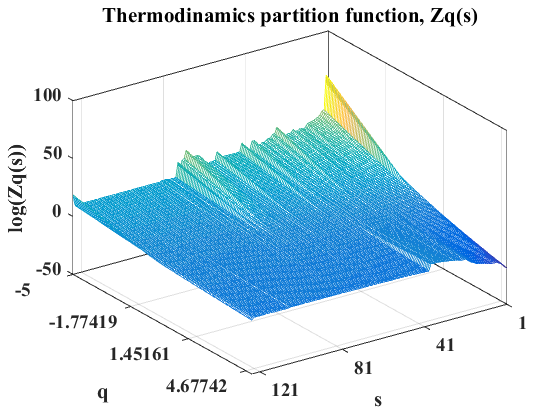} 
\includegraphics[scale=0.39]{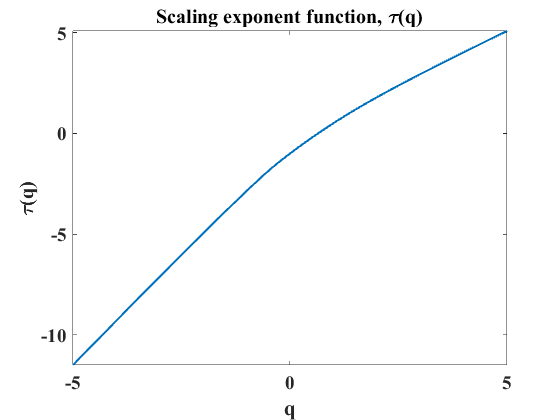}
\includegraphics[scale=0.39]{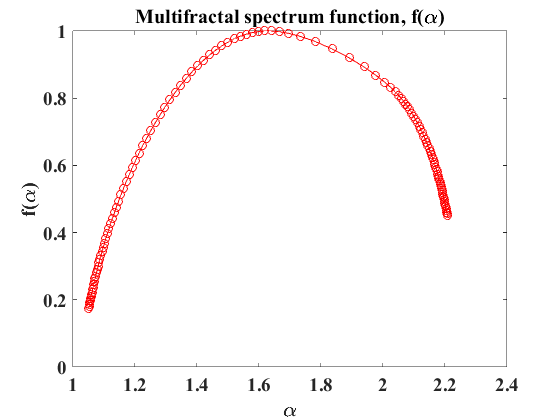}
\caption{The thermodynamic partition function $Zq$($s$) (left), the scaling exponent function $\tau$($q$) (middle), and the multifractal spectrum function $f$($\alpha$) (right) of image B of Q0957+561 in the $r$ band.}
\label{fig3}
\end{figure*}

\begin{figure*}
\centering
\includegraphics[scale=0.4]{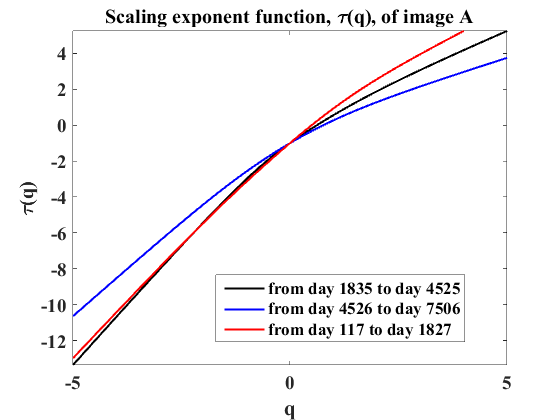} 
\includegraphics[scale=0.4]{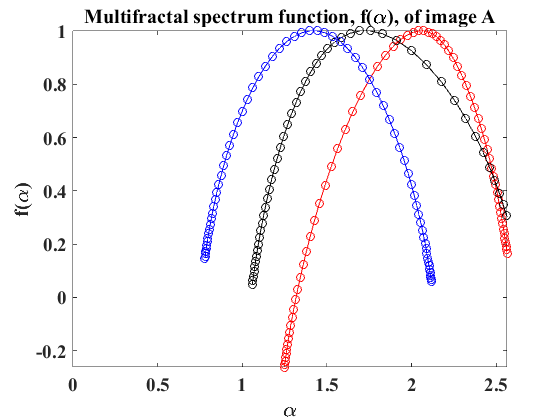}
\includegraphics[scale=0.4]{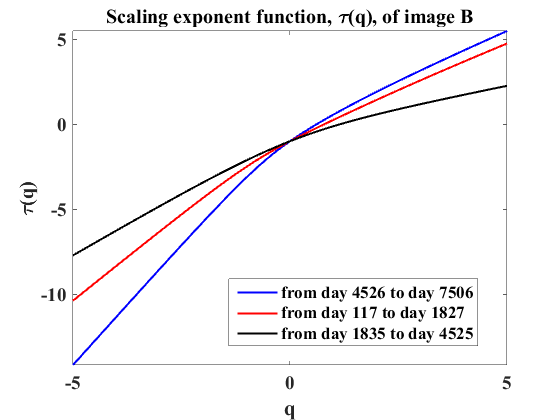}
\includegraphics[scale=0.4]{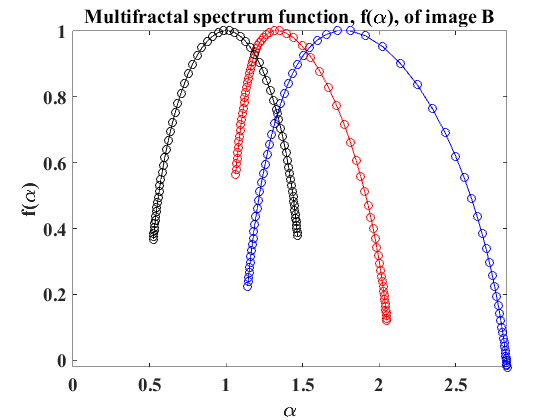} 
\caption{Top panel: The scaling exponent functions $\tau$($q$) (left) and the multifractal spectrum functions $f$($\alpha$) (right) for all time segments of image A of Q0957+561 in the $r$ band. Bottom panel:  The same as the upper panels, but for image B of Q0957+561 in the $r$ band. For both images: red (segment 1, from day 117 to day 1827), black (segment 2, from day 1835 to day 4525), and blue (segment 3, from day 4526 to day 7506).}
\label{fig4}
\end{figure*}

Using the skeleton functions, the collected local maxima lines, constructed for the two images in the $r$ band (top panel in Fig. \ref{fig1}) we determine the relationship between the logarithm of the thermodynamic partition function $Zq$($s$) and the scale $s$ for images A and B as shown in Figs. \ref{fig2} and \ref{fig3}, respectively. As one can see, the thermodynamic partition functions fluctuate in a nonlinear manner, revealing the nonlinear functionality between $Zq$($s$) and $s$. 
Though we have information about how $Zq$($s$) changes against $s$ at this level, it is usual to determine the scaling exponent functions $\tau$($q$), the slope of the log-log plots of the thermodynamic partition function $Zq$($s$) and the scale $s$, for both images to confirm the true relationship between the thermodynamic partition function and the scale. A nonlinear relationship between the thermodynamic partition function $Zq$($s$) and the scale $s$ observed, based on the thermodynamic partition function against the scale $s$ plots in (Figs. \ref{fig2} and \ref{fig3}), is further confirmed by the scaling exponent function $\tau$($q$) versus the moment $q$ plots for images A and B.
The nonlinearity between $\tau$($q$) and $q$, which is the slope of log($Zq$(s)) against log($s$), clearly indicates the presence of multifractal (nonlinear) structure in the light curves of the two images. From the scaling exponent functions $\tau$($q$) versus $q$ (Figs. \ref{fig2} and \ref{fig3}), the difference in the degree of nonlinearity between images A and B is visible. Once we are sure about the presence of multifractal (nonlinear) signature in the considered light curves based on the nonlinear scaling exponent function, the next step is determining the degree of multifractality or nonlinearity (i.e., how strong is the observed multifractal signature?) for the signals of both images. To that end, we estimate, and plot, the multifractal spectrum functions $f$($\alpha$) and calculate the width ($\Delta\alpha$ = $\alpha_{max}$ - $\alpha_{min}$) using equations Eqs. \ref{eq10} and \ref{eq11}. The calculated width $\Delta\alpha$ values for images A and B are $\Delta\alpha_{A}$ = 1.6030 and $\Delta\alpha_{B}$ = 1.1567, respectively. Wider width values confirm the presence of strong multifractal signatures and the intermittent nature of the light curves. The difference in nonlinearity observed between images A and B in the scaling exponent function plots, that the nonlinearity strength of image A is stronger than image B, is further confirmed by the calculated width values of the multifractal spectrum function (Figs. \ref{fig2} and \ref{fig3}). \\
To verify whether there is a change in the degree of multifractality in the long-term light curves  of images A and B in the $r$ band,  top panel in Fig. \ref{fig1}, we divide their light curves into three time segments containing equal data points: from day 117 to day 1827 (segment 1), from day 1835 to day 4525 (segment 2), and from day 4526 to day 7506 (segment 3). We perform multifractality analysis for each time segment separately. Following the same procedures applied in the previous subsections, we have computed absolute wavelet coefficients, constructed skeleton functions, and determined the thermodynamic partition functions $Zq$($s$) using the collected absolute wavelet coefficients that only hold local maxima lines. The slope calculated from the log-log plots of the thermodynamic partition function $Zq$($s$) against the scale $s$ is represented by the scaling exponent $\tau$($q$) (Fig. \ref{fig4}). In addition, the corresponding multifractal spectrum function $f$($\alpha$) is estimated (Fig. \ref{fig4}). The calculated width $\Delta\alpha$ for each time segment of the quasar images A and B are 1.3150/1.6293/1.3375 and 0.9888/0.9416/1.6943 (segment 1/segment 2/segment 3),  respectively. The nonlinear behaviour of the scaling exponent and the corresponding width values reveal the presence of strong multifractal (nonlinear) signature in each time segment of the light curves. Comparing the degree of multifractality (nonlinearity) between the corresponding light curves of the images, the multifractality (nonlinearity) strength of image A is found to be stronger than that of image B in the first two time segments, from day 117 to day 1827 and from day 1835 to day 4525. In contrast, image B is stronger in the last time segment (segment 3). For both images, A and B,  the order of time segments from the highest to the lowest in the degree of multifractality (nonlinearity) is segment 2, segment 3, and segment 1 (for image A) and segment 3, segment 1, and segment 2 (for image B). 
\begin{figure*}
\centering
\includegraphics[scale=0.4]{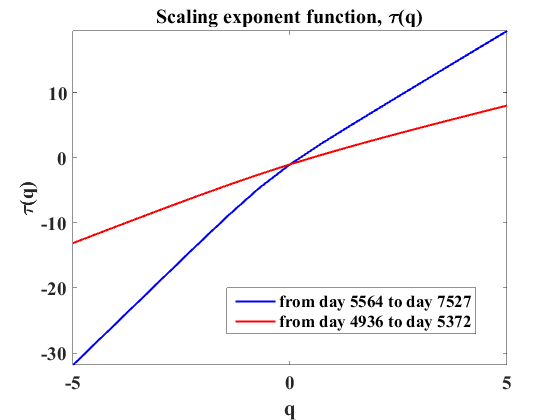}
\includegraphics[scale=0.4]{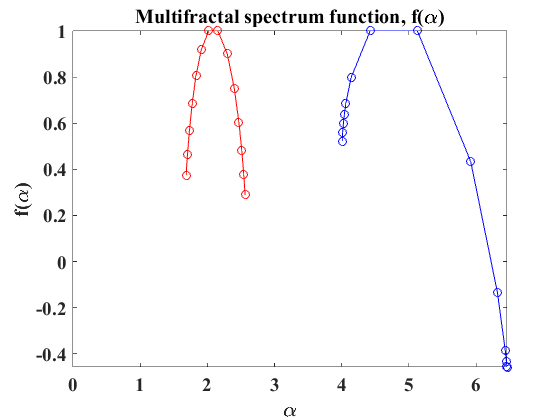} 
\caption{The scaling exponent functions $\tau$($q$) (left) and the corresponding multifractal spectrum functions $f$($\alpha$) (right) of image B in two different time segments, red (segment 1, from day 4936 to day 5372) and blue (segment 2, from day 5564 to day 7527).}
\label{fig5}
\end{figure*}
Though microlensing effects were detected in the initial years of the quasar monitoring \citep{1998A&A...336..829P},  strong evidence of microlensing was also detected in the last years of monitoring \citet{2018A&A...616A.118G}. Of the two quasar images, image B is most likely affected by microlensing since it is the closest to the centre of the galaxy and thus crosses an internal region of the galaxy's luminous halo. To investigate the effect of microlensing on image B, particularly in the last years of monitoring, we divide the light curve of the image (top panel in Fig. \ref{fig1}) from day 4936 to day 7527 into two equal time segments, though the number of data points is small -- from day 4936 to day 5372 (segment 1) and from day 5564 to day 7527 (segment 2) -- and perform the same multifractality analysis as in the previous cases. The results obtained -- namely, the scaling exponent and multifractality spectrum function, which are given in Fig. \ref{fig5} -- clearly show the existence of multifractal behaviour. The calculated width values are 0.8791 and 2.4548 in segment 1 and segment 2, respectively. The degree of multifractality detected in segment 2 is stronger than the one detected in segment 1. Considering the results obtained here for segment 2 and the one obtained in the previous analysis for segment 3 (from day 4526 to day 7506), we can see that the degree of multifractality of parts of the light curve in image B (specifically, parts that include data points in the range of day 5564 to day 7500) is stronger than those that did not include this range.
\subsection{Analysis of the light curves of image A in the $r$ and $g$ bands}\label{subsec2}
\begin{figure*}
\centering
\includegraphics[scale=0.39]{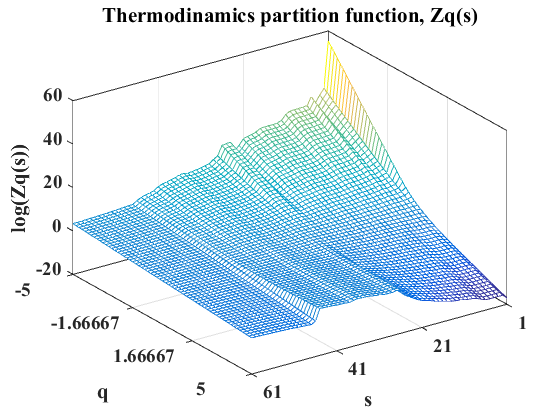} 
\includegraphics[scale=0.39]{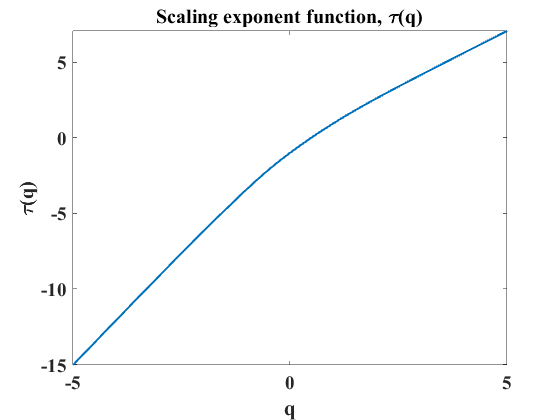}
\includegraphics[scale=0.39]{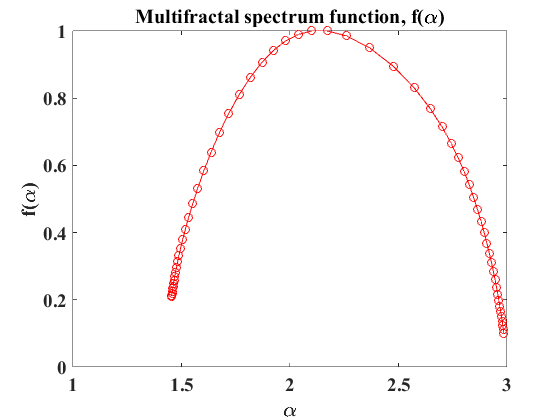}
\includegraphics[scale=0.39]{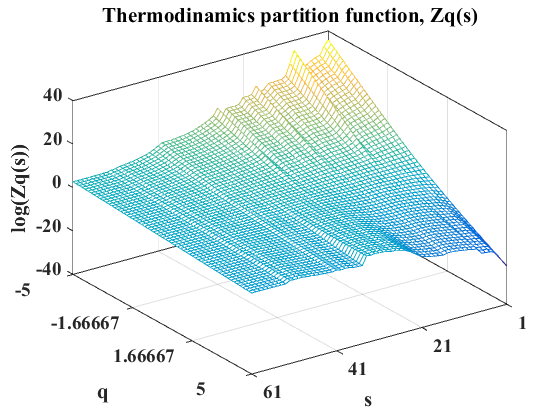} 
\includegraphics[scale=0.39]{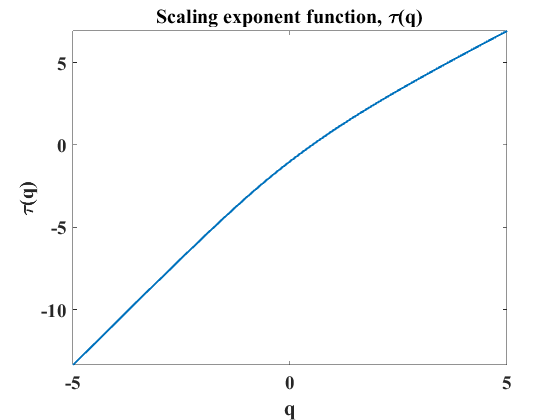}
\includegraphics[scale=0.39]{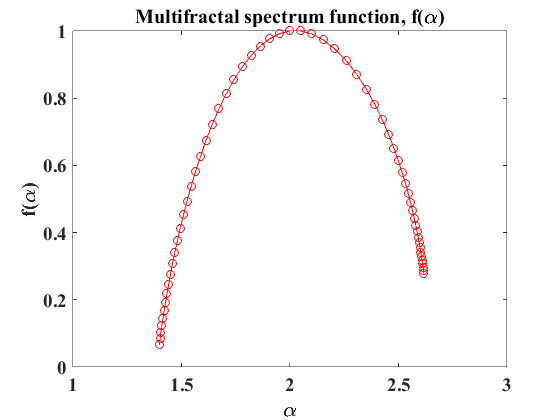}
\caption{Upper panel: The thermodynamic partition function $Zq$($s$) (left), the scaling exponent function $\tau$($q$) (middle), and the multifractal spectrum function $f$($\alpha$) (right) of image A of Q0957+561 in the $r$ band. Bottom panel: The same as the upper panels, but for image A of Q0957+561 in the $g$ band.}
\label{fig6}
\end{figure*}

Here, we analysed the light curve of image A in the $r$ and $g$ bands (Fig. \ref{fig1}, middle panel). By using the collected local maxima lines (the skeleton functions) for the light curves, we determine the corresponding thermodynamic partition functions $Zq$($s$) against the scale $s$ as presented in Fig. \ref{fig6}. The slope of the log-log plots of the thermodynamic partition functions $Zq$($s$) versus the scale $s$ in the $r$ and $g$ bands is represented by the scaling exponent $\tau$($q$) in Fig. \ref{fig6}, revealing the nonlinear relationship between $Zq$($s$) and $s$. 
To further confirm the observed nonlinear relationship between the thermodynamic partition function $Zq$($s$) and the scale $s$, and to determine the strength of the detected multifractal (nonlinear) signature, we estimate the multifractal spectrum functions for the signal of image A in the $r$ and $g$ bands. The estimated multifractal spectrum functions $f$($\alpha$) of image A in the $r$ and $g$ bands are shown in Fig. \ref{fig6}. The calculated width values in $r$ and $g$ bands are $\Delta\alpha_{A_{r}}$ = 1.5319 and $\Delta\alpha_{A_{g}}$ = 1.22, respectively, proving the multifractal behaviour of image A of Q0957+561 in the $r$ and $g$ bands. Here also, there is a clear difference in the degree of nonlinearity between intrabands: the nonlinearity in the $r$ band is observed to be stronger than the one detected in the $g$ band. 
\subsection{Analysis of the light curves of image B in the $r$ and $g$ bands}\label{subsec3}
Similarly, following the same procedures as discussed in the last subsections, we have analysed the light curves of image B of Q0957+561 in the $r$ and $g$ bands ( Fig. \ref{fig1}, bottom panel).
\begin{figure*}
\centering
\includegraphics[scale=0.39]{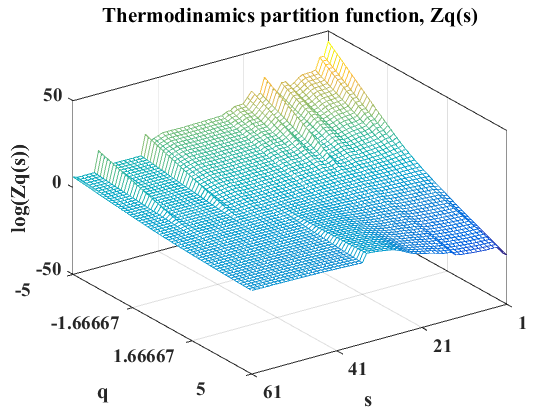} 
\includegraphics[scale=0.39]{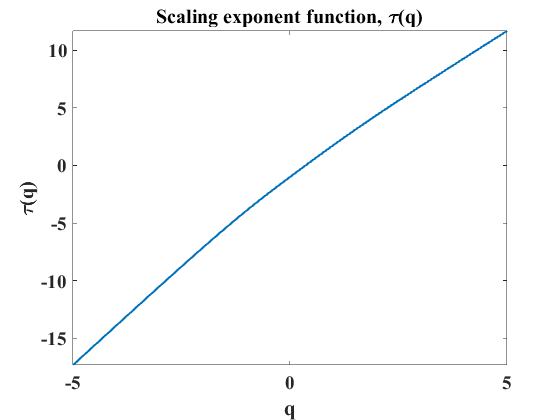}
\includegraphics[scale=0.39]{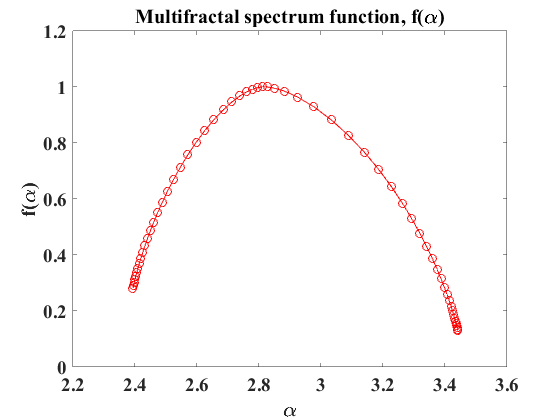}
\includegraphics[scale=0.39]{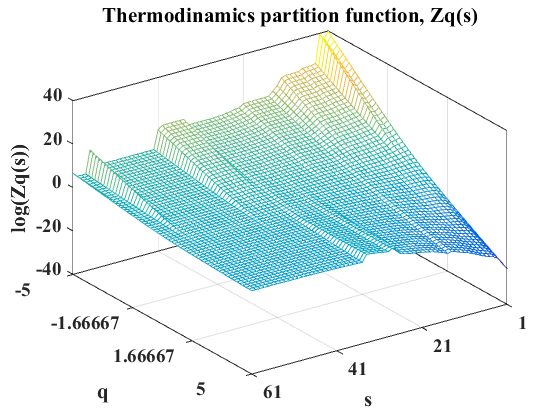} 
\includegraphics[scale=0.39]{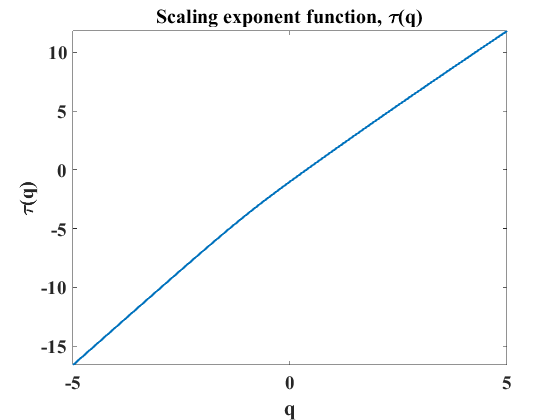}
\includegraphics[scale=0.39]{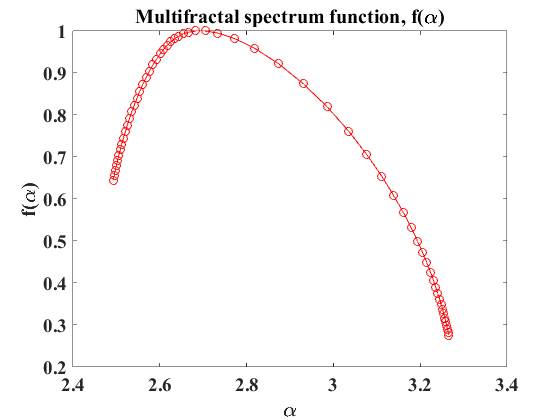}
\caption{Upper panel: The thermodynamic partition function $Zq$($s$) (left), the scaling exponent function $\tau$($q$) (middle), and the multifractal spectrum function $f$($\alpha$) (right) of image B of Q0957+561 in the $r$ band. Bottom panel: The same as the upper panels, but for image B of Q0957+561 in the g band.}
\label{fig7}
\end{figure*}
The thermodynamic partition functions $Zq$($s$) calculated by using the skeleton functions for the light curves of image B in the $r$ and $g$ bands, which in turn were constructed from the absolute wavelet coefficients of the corresponding light curves, are given in Fig. \ref{fig7}. The scaling exponent functions $\tau$($q$), i.e., the slope of the log-log plot of $Zq$($s$) against $s$, is also presented in Fig. \ref{fig7}. The $\tau$($q$) versus $q$ plot clearly shows the presence of nonlinear signatures in the light curves of image B in the $r$ and $g$ bands. The nonlinearity or multifractality detected in the light curves is further strengthened by the estimated multifractal spectrum function in Fig. \ref{fig7}. The calculated width ($\Delta\alpha$ = $\alpha_{max}$ - $\alpha_{min}$) values are $\Delta\alpha_{B_{r}}$ = 1.0481 and $\Delta\alpha_{B_{g}}$ = 0.7739 for the $r$ and $g$ bands of image B of Q0957+561, respectively, confirming the conclusion based on the scaling exponent function. As discussed in subsections \ref{subsec2} and \ref{subsec3}, for both quasar images, the degree of multifractality is stronger in the $r$ band.
\section{Summary and Conclusions}\label{concl}
We analyse the presence of multifractal (nonlinear) signatures in the light curves of images A and B of the lensed quasar Q0957+561 in the optical $r$ and $g$ bands using a wavelet transform modulus maxima-based multifractality analysis technique. In addition, we divide the long-term monitoring light curves of the quasar images in the $r$ band into different time segments and repeat the same multiractality analysis for each time segment separately. First, we calculate the absolute wavelet coefficients using the continuous wavelet transform approach and form a matrix of local maxima lines (construct skeleton functions) by aggregating the absolute wavelet coefficients that only hold maxima lines. Second, using the constructed skeleton function, we determine the thermodynamics partition function for all the light curves considered. Third, we estimate the slope of the log-log plots of the thermodynamic partition function $Zq(s)$ and the scale $s$. The estimated behaviours of the slopes are quantified by the scaling exponent function $\tau$($q$) versus the moment $q$ plots. Finally, we estimate the multifractal spectrum at each frequency for all the light curves and calculate the degree of multifractality from the width $\Delta\alpha$ of the spectrum. Our main findings are the following: (i) we observed strong multifractal behaviour in all the light curves analysed; (ii) the degree of multifractality for both images in the $r$ band changes over time in a non-monotonic way; (iii) in the r band, in periods of quiescent microlensing activity, we found that the degree of multifractality (nonlinearity) of image A is stronger than that of B, while B has the larger multifractal strength in recent epochs (from day 5564 to 7527), when it appears to be affected by microlensing; and  (iv) in a period of quiescent microlensing activity in the $g$ and $r$ bands, the degree of multifractality is stronger in the $r$ band for both quasar images.\\
The detection of a multifractal signature in the quasar light curves is in agreement with our previous results that quasars are intrinsically multifractal or nonlinear systems \citep{2018Belete, doi:10.1093/mnras/sty1316}. The observed multifractality could be due to different physical mechanisms. It is the variation in flux that results in a multifractal signature in a light curve. It has been identified that reverberation within the gas disc around the supermassive black hole is responsible for most observed variations in Q0957+561 \citep{shalyapin2008new}; therefore, most likely it is this physical mechanism that causes multifractal (nonlinearity) signatures in the light curves considered. Though no extrinsic signals or microlensing effects  were detected for decades in the light curves of the quasar Q0957+571 \citep{shalyapin2008new}, recently, strong evidence for the presence of microlensing effects in the light-curves of Q0957+561 has been found \citep{2018A&A...616A.118G}. It is most likely that quasar images in the same band, in the optical in our case, have similar (if not the same) radiation mechanisms and regions and are consequently expected to have similar nonlinear behaviours unless otherwise contaminated in a way that changes the intrinsically nonlinear structure. This effect is because for intrinsically variable quasars, the fluxes measured from the images are expected to have similar light curves, except for certain lags -- time delays -- and an overall offset in magnitude \citep{wambsganss1998gravitational}; as a result, the  light curves are expected to be similarly nonlinear or multifractal. 
If there were no contribution from extrinsic variation, mainly due to microlensing, the nonlinear signature of the signals would not differ that much. Also it has been indicated, at a given epoch, intrinsic variability should affect images of a lensed quasar in the same way, whereas microlensing may induce differences between the spectra of different images. Therefore, assuming that all signals of the images in the $r$ band have similar (if not the same) radiation mechanisms and regions (from the accretion disk or continuum compact source), or both signals are intrinsically similarly nonlinear, any difference in their nonlinearity strength would most likely be due to the existence of extrinsic variabilities of a different nature in the observed light curves of the images or due to microlensing by stars in the lensing galaxies affecting image B, since microlensing affects the light curves of quasar images \citep{2018MNRAS.476..663K}. Therefore, the change in the degree of multifractality over time in a non-monotonic way provides us physically important information about the existence of extrinsic variations in the observed light curves of the quasar images. In other words, the observed difference in the degree of multifractality between the time segments of the light curves indicates that the images are affected by different physical phenomena along their paths to the observer, and consequently, the degree of multifractality is different between them. In particular, in the case of image B, the increase in the degree of multifractality in the last years of monitoring could be taken as an additional evidence for the presence of extrinsic variability due to microlensing effects since it is the one presumably playing a more relevant role in the image B. The nonlinearity in both quasar images (leaving aside microlensing, in addition to the "excesses" in A and $r$) should have an intrinsic origin (within the source quasar). The observed nonlinearity may be produced in the central high-energy source that irradiates the accretion disk (AD), when its central flares are reprocessed in the AD and/or when these flux variations from the AD are again reprocessed in the broad-line region (BLR). Moreover, in the absence of microlensing, both observed "excesses" of nonlinearity are most likely associated with the presence of a compact dusty region in the lensing galaxy.  However, it remains unclear how dust extinction works to generate these "excesses".
A significant "excess" of nonlinearity is generated when the BLR reprocesses the radiation from the compact sources. The resulting diffuse light plays a more important role in A because its direct light is partially extinguished by the compact dusty cloud in the lensing galaxy. This rationale can explain the "excess" in A (in relation to B). The diffuse contribution could also be more important in the $r$ band (in relation to $g$), which would explain the "excess" in $r$ in both images. It has been indicated that measuring the abundance and strength of nonlinear potential fluctuations along sightlines to high redshift provides a powerful test of cosmic structure formation scenarios \citep{cen1994strong}. We believe that our results provide significant information to better understand the physical properties of the intervening medium and construct a model to better understand the matter distribution in foreground lensing galaxies. Furthermore, multifractality (nonlinearity) analysis could be used to check whether signals from background sources are only intrinsic or a combination of both intrinsic and extrinsic signals, in addition to techniques already developed to do so. To the best of our knowledge, this is the first time that this nonlinearity analysis technique has been applied to extragalactic lensed sources, and we consider that this approach could be used to study the relationship between the change in the nonlinear structure of intrinsic signals from different background sources at different redshifts, and mass, of lensing galaxies based on the change in the multifractal (nonlinear) behaviour of the signals, from which we can learn much about the nature of the matter distribution in lensing galaxies and background quasi-stellar objects. 
\section*{Acknowledgements}
This publication makes use of data from the GLQ database in the GLENDAMA archive available at  http://grupos.unican.es \\ /glendama/database/. The research activities of the Observational Astronomy Board of the Federal University of Rio Grande do Norte (UFRN) are supported by continuous grants from CNPq, CAPES and FAPERN Brazilian agencies. We also acknowledge financial support from INCT INEspa\c{c}o/CNPq/MCT. AB acknowledges a CAPES PhD fellowship. ICL acknowledges a CNPq/PDE fellowship. We warmly thank the anonymous reviewer for the fruitful comments and suggestions that greatly improved this work.

\nocite{*}
\bibliographystyle{mnras}
\bibliography{biblio} 



\label{lastpage}
\end{document}